\documentclass[twocolumn,pra,amsmath,amssymb,showpacs,superscriptaddress]{revtex4-1}
\usepackage{epsfig}
\usepackage[dvipsnames]{xcolor}
\usepackage{color}
\usepackage{graphicx,color}
\usepackage{dcolumn}
\usepackage{bm}
\usepackage{braket}
\usepackage{amsthm,amssymb,bm}
\usepackage[colorlinks=true,
           linkcolor=red,
            urlcolor=blue,
           citecolor=ForestGreen]{hyperref}
\usepackage{array}

\usepackage[normalem]{ulem}

\providecommand{\openone}{\leavevmode\hbox{\small1\kern-3.8pt\normalsize1}}

\begin{document}
\title{Dissipative dynamics and cooling rates of trapped impurity atoms immersed in a reservoir gas}

\author{R. G. Lena}
\affiliation{Department of Physics, SUPA and University of Strathclyde, Glasgow G4 0NG, United Kingdom}
\author{A. J. Daley}
\affiliation{Department of Physics, SUPA and University of Strathclyde, Glasgow G4 0NG, United Kingdom}

\date{\today }

\begin{abstract}
We study the dissipative dynamics of neutral atoms in anisotropic harmonic potentials, immersed in a reservoir species that is not trapped by the harmonic potential. Considering initial motional excitation of the atoms along one direction, we explore the resulting spontaneous emission of reservoir excitations, across a range of trap parameters from strong to weak radial confinement. In different limits these processes are useful as a basis for analogies to laser cooling, or as a means to introduce controlled dissipation to many-body dynamics. For realistic experimental parameters, we analyse the distribution of the atoms during the decay and determine the effects of heating arising from a finite temperature reservoir.
\end{abstract}


\maketitle

\section{Introduction}

Developments in experiments with ultracold atoms over the past two decades have allowed not only the exploration of coherent many-body dynamics, but also dissipative dynamics, in limits that are well-understood based on microscopic considerations \cite{Muller2012, Daley2014b}. This has enabled both additional control over unwanted dissipation in these systems \cite{Ashkin1980, Castin1998, Wolf2000, Gerbier2010, Pichler2010, McKay2011}, and opportunities to explore the effects of dissipation on many-body systems, including robustness to decoherence \cite{Schreiber2015, Levi2016, Luschen2017, ospelkaus2006localization} and dissipative driving as a means to prepare desired many-body states \cite{Diehl2008, Kraus2008, Verstraete2009, Yi2012}.

In this context, the study of dissipation induced by immersing the system in a reservoir of a different species is particularly intriguing, offering a means to cool atoms to a motional ground state without destroying internal states under appropriate conditions \cite{DaleyPRA}. For large reservoirs, this process takes the form of spontaneous emission of a reservoir excitation, with the decay of the atom being mathematically analogous in many ways to the decay of an excited atom via spontaneous emission of photons. This has a variety of potential applications, including implementation of dark-state laser cooling schemes for reducing the temperature of atoms in an optical lattice within a band [\cite{Griessner2006, Griessner2007}]. Initial experiments have demonstrated cooling from higher bands \cite{Scelle2013,Chen2014}, but were often limited by collisions between atoms in higher bands. The development of dual-species experiments with Alkaline-Earth-Metal atoms and Alkali atoms \cite{Tey2010,vaidya2015degenerate,spethmann2012dynamics} now provide new opportunities in this direction. Especially in the case of spin-polarised fermions, where collisions within the lattice are suppressed, the dynamics should be dominated by the coherent dynamics of the system and dissipation induced by coupling of atoms between bands. 

Inspired by these ongoing opportunities, in this article we quantitatively analyse these dissipative processes for atoms in anisotropic traps. Previous theoretical studies \cite{DaleyPRA, Griessner2007} have generally relied on 1D models for the trapped atoms, assuming strong confinement in the radial directions. Here we consider a range of trapping conditions in the radial direction with respect to the direction along which an atom is initially excited. This allows us also to treat parameters where the radial trapping is weak. This regime is both a natural starting point for experiments (with lattices created in 1D or 2D), and provides the intriguing possibility of cooling distributions of fermions on a single lattice site. It is also a natural starting point for considering dissipative transport dynamics, which have recently been considered between harmonic traps, in which the radial states served as a continuum of final states for an effective dissipative process in atomtronics \cite{Seaman2007a,Labouvie2015}. The parameter regimes considered here would provide a new way to further control such dissipative transport dynamics.

We first study the spontaneous emission of reservoir excitations from a single impurity atom initially excited along the axial direction, deriving the corresponding master equation and evaluating the transition coefficients between axial and radial states. We study the dependence of these rates on the frequencies, and determine realistic decay times based on parameters used in current experiments. We consider both the case of strong radial trapping, where the effect of the radial frequency is primarily quantitative, and then the case of weak radial trapping, where the physics qualitatively changes, as described above. We then consider both the effects of finite temperatures, and the dynamics of fermions with weak radial trapping.

The remainder of the manuscript is organised as follows: In Sec.~\ref{Sec:Model} we introduce the model and derive the master equation of the open system under the Born-Markov approximation, giving an overview of preliminary calculations and concepts used in the following sections. In Sec.~\ref{Sec:3D} we then study the spontaneous emission of an impurity in a 3D harmonic trap tightly confined in one direction and isotropic in the other two, and see how the dynamics of the atom initially excited changes when varying the ratio between the trapping frequencies. In Sec.~\ref{Sec:2D} we consider in more detail the case of weak trapping in the radial direction and analyse the effects of heating due to non-zero reservoir temperatures. We then generalise the results to the case of spin-polarised fermions in Sec.~\ref{Sec:fermions}, before discussing the conclusions and outlook in Sec.~\ref{Sec:conclusions}.


\section{Model}
\label{Sec:Model}
We consider an impurity atom in a harmonic trap immersed in a 3D superfluid reservoir, where for simplicity we consider the latter to be confined in a well potential of volume $V$ and we neglect the internal degrees of freedom of the impurity.
The model is described by the total Hamiltonian
\begin{equation}
H = H_{a}+H_{b}+H_{int},
\end{equation}
where
\begin{equation}
H_a = \hbar \left( \omega_x \hat{n}_x+ \omega_y \hat{n}_y + \omega_z \hat{n}_z+\frac{1}{2}(\omega_x+\omega_y+\omega_x)\right),
\end{equation}
is the Hamiltonian for the impurity, described by a 3D quantum harmonic oscillator,
\begin{equation}
H_{b} = E_0+\sum_{\textbf{k}\neq 0}\epsilon(\textbf{k})\hat{b}^{\dagger}_\textbf{k}\hat{b}_\textbf{k},
\end{equation}
is the Hamiltonian of the superfluid bath, obtained from Bogoliubov theory of a weakly interacting Bose gas \cite{pethick2008bose}, so that $\hat{b}^{\dagger}_\textbf{k}$ and $\hat{b}_\textbf{k}$ create and annihilate Bogoliubov excitations with energy $\epsilon(\textbf{k})=\epsilon_k$ and momentum $\hbar \textbf{k}$, and where $E_0$ is the ground state energy of the superfluid.

The contact interaction between the system and reservoir is given by the Hamiltonian of the form
\begin{align}
H_{int}&= g_{ab}\int \delta \hat{\rho}(\textbf{r}_b)\delta(\textbf{r}-\textbf{r}_b)d^3\textbf{r}_b=g_{ab}\delta \hat{\rho}(\textbf{r}),
\end{align}
where $\textbf{r}$ is the position operator for the motional states of the impurity and where $\textbf{r}_b$ and $\delta\hat{\rho}$ are respectively the position and the density fluctuation operators of the superfluid. 
The coupling strength between the impurity of mass $m_a$ and the atoms of the BEC (with mass $m_b$) is given by $g_{ab}=4\pi \hbar^2 a_{ab}/2\tilde{m}$, where $a_{ab}$ is the scattering length between the impurity and the superfluid, and $\tilde{m}=\frac{m_a m_b}{m_a+m_b}$ is the reduced mass.
The density fluctuation operator $\delta\hat{\rho}$ was obtained by using a mean field description for the field operator $\hat{\Psi}=\sqrt{\rho_0}+\delta \hat{\Psi}$ (see Appendix ~\ref{Appendix_master_equation}).
Under the assumption of a weakly interacting Bose gas at low temperatures, the terms at the second order in $\delta\hat{\Psi}$ can be neglected, and by noting $\delta\hat{\Psi}=\frac{1}{\sqrt{V}}\sum_{\textbf{k}}(u_{\textbf{k}}\hat{b}_{\textbf{k}}e^{i\textbf{k}\cdot\textbf{r}}+v_{\textbf{k}}\hat{b}_{\textbf{k}}^{\dagger}e^{-i\textbf{k}\cdot\textbf{r}})$, the interaction Hamiltonian reduces to
\begin{align}
H_{int}&=g_{ab}\sqrt{\rho_0}(\delta\hat{\Psi}^{\dagger}(\hat{r})+\delta\hat{\Psi}(\hat{r}))\nonumber\\
&=g_{ab}\sqrt{\frac{\rho_0}{V}}\sum_\textbf{k}[(u_\textbf{k}+v_\textbf{k})(\hat{b}_\textbf{k}e^{i\textbf{k}\cdot\hat{\textbf{r}}}+\hat{b}^{\dagger}_\textbf{k}e^{-i\textbf{k}\cdot\hat{\textbf{r}}})],
\label{eq:H_int}
\end{align}
where $u_{\textbf{k}}$ and $v_{\textbf{k}}$ are the coefficients obtained from the Bogoliubov transformation of the form
\begin{align}
u_{\textbf{k}}^2&=\dfrac{R_{\textbf{k}}^2}{1-R_{\textbf{k}}^2}\\
v_{\textbf{k}}^2&=\dfrac{1}{1-R_{\textbf{k}}^2},
\end{align}
having defined $R_{\textbf{k}}=\dfrac{\epsilon_{\textbf{k}}-\epsilon_{\textbf{k}}^{(sup)}-\mu}{\mu}$, where $\mu=g_{bb}\rho_0=m_bu^2$ is the chemical potential of the reservoir, with $g_{bb}$ the boson-boson interaction strength, $\rho_0$ is the density of the BEC and $u=\sqrt{\dfrac{g_{bb}\rho_0}{m_b}}$ is the speed of sound in the superfluid. The energy of the excitations is given by
\begin{equation}
\epsilon_{\textbf{k}}=\sqrt{\left(\epsilon_\textbf{k}^{(sup)}\right)^2+\left(\epsilon_{\textbf{k}}^{(sub)}\right)^2},
\end{equation}
with $\epsilon_\textbf{k}^{(sup)}=\dfrac{\hbar^2k^2}{2m_b}$ and $\epsilon_\textbf{k}^{(sub)}=\hbar u k$ the energy of the Bogoliubov excitations respectively in the supersonic regime (when $\epsilon_k\gg\mu$, up to the chemical potential) and in the subsonic regime (for $\epsilon_k\ll\mu$).
The different dispersion relations in the two regimes imply a change also in the structure factor $S(\textbf{k})=(u_{\textbf{k}}+v_{\textbf{k}})^2$, with $S(\textbf{k})\sim 1$ in the supersonic limit and $S(\textbf{k})\simeq \dfrac{\hbar k}{2 m_b u}$ in the subsonic limit.

\begin{figure}[t]
\includegraphics[scale=.6]{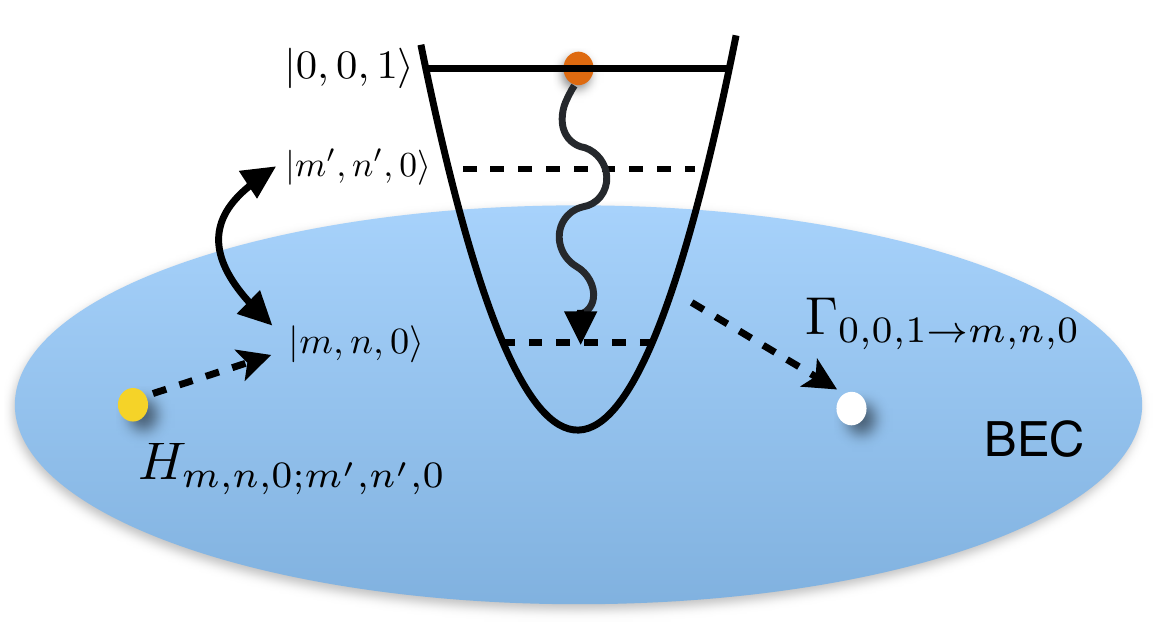} 
\caption{Schematic representation of level transition mechanisms for an atom initially excited along one direction in a 3-D harmonic trap immersed in
a superfluid. At $T=0$ the only possible transitions are given by the decay from the state $\ket{0,0,1}$ to the state $\ket{m,n,0}$ with the creation of Bogoliubov excitations and are described by the coefficients $\Gamma_{001\rightarrow mn0}$ in the equations of motion.
At finite temperature, an additional contribution due to the interaction with thermal excitations can induce stimulated transitions and excite the atoms to higher motional states either radially or axially. This contribution is represented by the coefficients $H_{m,n,\alpha;m',n',\alpha'}$.}
\label{fig:transitions}
\end{figure}

After deriving the master equation under the Born-Markov approximation (see Appendix~\ref{Appendix_master_equation} for further details) we find the occupation probability of the impurity in the state $\ket{m_x,m_y,m_z}$ to be given by
\begin{align}
\dot{p}_{m_{x,y,z}}=&\sum_{\substack{n_{x,y,z}:\\ \sum_i \omega_i(n_i-m_i) > 0}} \Gamma_{n_{x,y,z}\rightarrow{m_{x,y,z}}} p_{n_{x,y,z}}\nonumber\\
& -\sum_{\substack{m'_{x,y,z}:\\ \sum_i \omega_i(m_i-m'_i)>0}} \Gamma_{m_{x,y,z}\rightarrow{m'_{x,y,z}}}p_{m_{x,y,z}}\nonumber\\
& + \sum_{n_{x,y,z}}H_{n_{x,y,z};m_{x,y,z}}(p_{n_{x,y,z}}-p_{m_{x,y,z}}).
\label{eq:occ_prob_3D}
\end{align}
As illustrated in Fig.~\ref{fig:transitions}, the first two terms in Eq.~\eqref{eq:occ_prob_3D} define the decay with the emission of a Bogoliubov excitation, while the third term describes stimulated emission and absorption of thermal excitations that can bring the atom to higher motional states.
These transition rates are derived according to Fermi's golden rule, as
 \begin{align}
\Gamma_{n_{x,y,z}\rightarrow m_{x,y,z}}&=\frac{2\pi}{\hbar}\sum_{\textbf{k}}\bigl|T_{n,m}(\textbf{k})\bigr|^2\delta(\tilde{\epsilon}-\epsilon_\textbf{k}),
\label{eq:decay_rate_general}\\
H_{n_{x,y,z};m_{x,y,z}}&=\frac{2\pi}{\hbar}\sum_{\textbf{k}}N(\textbf{k})\bigl|T_{n,m}(\textbf{k})\bigr|^2\delta(\tilde{\epsilon}-\epsilon_\textbf{k}),
\label{eq:stimulated_transition_rate_general}
\end{align}
where $\tilde{\epsilon}$ is the difference of energy between initial and final state of the impurity,
\begin{align}
T_{n,m}(\textbf{k})=g_{ab}\sqrt{\frac{\rho_0}{V}}(u_\textbf{k}+v_\textbf{k})\prod_{i=x,y,z}\braket{m_i|e^{-ik_i r_i}|n_i},
\label{eq:matrix_elements}
\end{align}
and $N(\textbf{k})=(e^{\beta\epsilon_\textbf{k}})^{-1}$ is the number of excitations with momentum $\textbf{k}$ given by the Bose distribution, to be taken into account when considering a finite temperature reservoir.

The effect of the thermal excitations can, in principle, be neglected if $k_BT_B\ll\hbar\omega$ for all relevant trapping frequencies $\omega$, but depending on the geometry of the system, this condition may not be fulfilled. In Sec.~\ref{Sec:2D} we study how heating effects modify the dynamics of the system and under which conditions they can be minimized and neglected.

In the following sections we focus on the derivation of these transition coefficients to study the evolution of the state of the impurity in different geometrical confinements, in the case of impurities initially excited in the first excited state along the tightly confined direction.

\section{Single atom cooling in a 3-D harmonic potential tightly confined in one direction}
\label{Sec:3D}
In this section we study the case of an impurity trapped in a 3D harmonic potential tightly confined in the axial direction $z$ and isotropic in the other directions, so that $\omega_z\gg\omega_r=\omega_{x,y}$. For this case of a pancake shaped potential, we will refer to the tightly confined direction as the axial direction, and to the others as the radial directions. 
In this scenario, we consider the atom initially in the first excited state along the axial direction, and we study the spontaneous emission with decay towards the radial directions.

At this aim, we restrict our study to the case where the separation between the energy levels in the different directions is much larger than the chemical potential of the BEC, so we can study the dynamics of the system in the supersonic regime, where the Bogoliubov excitations, emitted during the decay of the excited impurities, are particle-like having energy $\epsilon_{\textbf{k}}=\hbar^2k^2/(2m_b)$ and the structure factor is $S(\textbf{k})=|u_{\textbf{k}}^2+v_{\textbf{k}}^2|\simeq 1$.

The spontaneous decay rates can be evaluated using Eq.~\eqref{eq:decay_rate_general} and Eq.~\eqref{eq:matrix_elements}, noticing that the term $e^{-i k_j \hat{r}_j}$ is the displacement operator $\hat{D}(\alpha)=\exp[\alpha\hat{a}_j^{\dagger}-\alpha^*\hat{a}_j]$, with $\alpha=-\dfrac{i k_j r_{j_0}}{\sqrt{2}}$ and $r_{j_0}=\sqrt{\dfrac{\hbar}{m_a\omega_j}}$ the oscillation length in the j direction. By using the identity \cite{barnett2002methods}
\begin{align}
\braket{n'|D(\alpha)|n}=\sqrt{\frac{n_>!}{n_<!}}e^{-\dfrac{|\alpha|^2}{2}}\alpha^{|n-n'|}L_{n_<}^{|n-n'|}(|\alpha|^2),
\label{eq:tr_coeff_Laguerre}
\end{align}
we then obtain
\begin{align}
\braket{n'_j|e^{-i k_j j_0}|n_j}=&\sqrt{\frac{n_<!}{n_>!}}e^{-\dfrac{r_{j_0}^2k_j^2}{4}}\biggl(-\dfrac{i j_0 k_j}{\sqrt{2}}\biggr)^{|n_j-n'_j|}\nonumber\\
&\times L_{n_<}^{|n_j-n'_j|}\biggl(\dfrac{r_{j_0}^2k_j^2}{2}\biggr),
\end{align}
with $n_<=\min(n_j,n_j')$, $n_>=\max(n_j,n_j')$, and $L_n^{(n_j-n_j')}(x)$ is the associated Laguerre polynomial.

After transforming the components of the momentum along the three directions to spherical coordinates and integrating over the momentum using the property of the delta function, we obtain that the dimensionless decay rates of Eq.~\eqref{eq:decay_rate_general} for the transitions $\ket{0_x,0_y,1_z}\rightarrow\ket{m_x,n_y,0_z}$ are given by
\begin{align}
\dfrac{\Gamma_{001\rightarrow mn0}}{\sqrt{\omega_r\omega_0}}=&\frac{2g_{ab}^2\rho_0\sqrt{m_a m_b}}{(2\pi)^2\hbar^3 u m!n! w}\left(\frac{m_b}{m_a}(w-(m+n))\right)^{m+n+\frac{3}{2}}\nonumber\\
&\times A_{\phi}(m,n)\int_0^{\pi}d\theta \cos^2\theta(\sin^2\theta)^{m+n+1/2}\nonumber\\
&\times e^{-\frac{m_b}{m_a}(w-(m+n))\left(\sin^2\theta+\frac{1}{w}\cos^2\theta\right)},
\label{eq:F001_mno}
\end{align}
where $A_{\phi}(m,n)=\int_0^{2\pi}d\phi\cos^{2m}\phi\sin^{2n}\phi$. We wrote them in units of $\sqrt{\omega_r \omega_0}$, with $\omega_0=\mu_b/2\hbar$, so that they would explicitly depend on the ratio $w=\omega_z/\omega_r$.

\begin{figure}[tb]
\centering
\includegraphics[scale=.42]{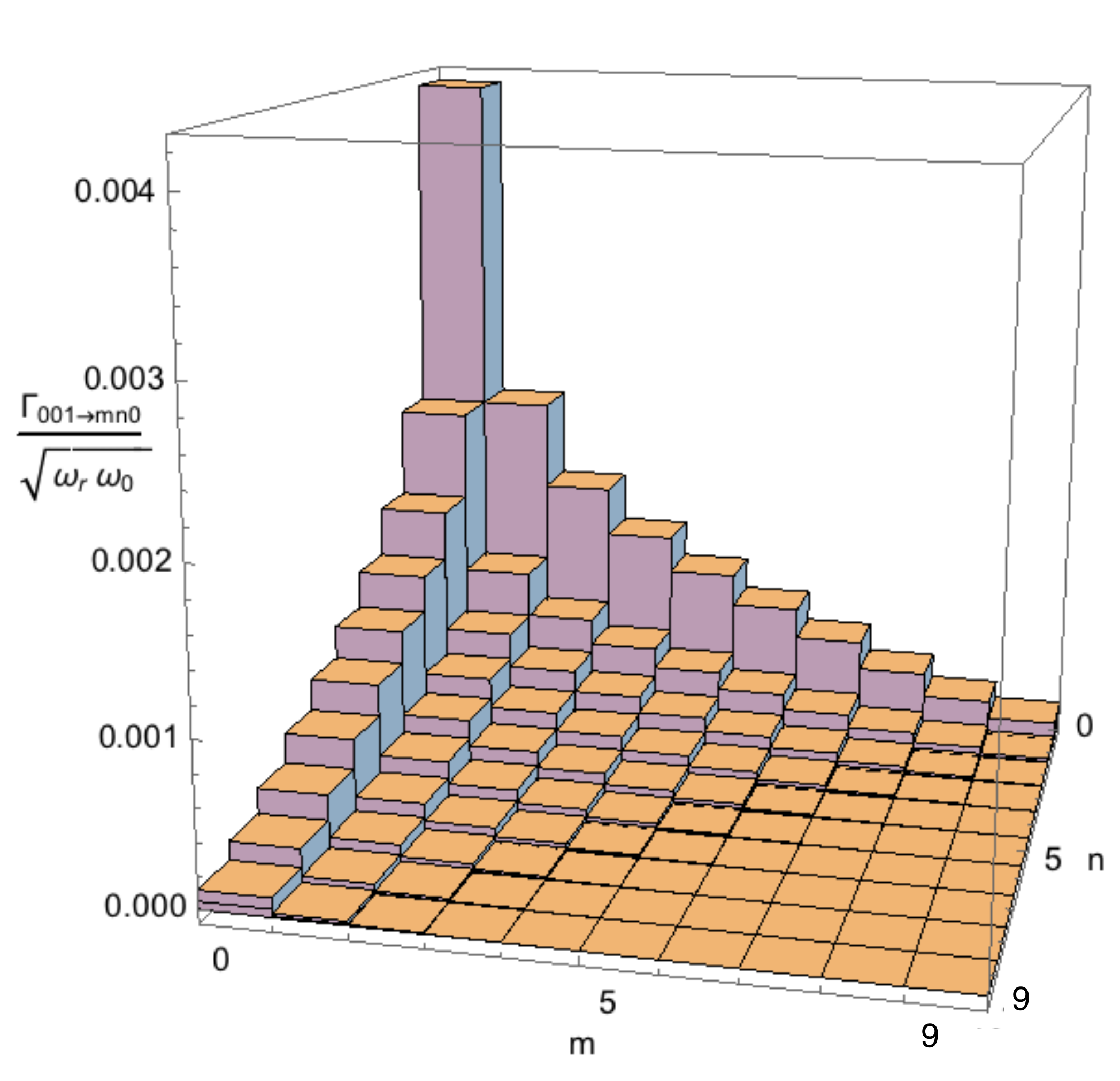} 
 \caption{Transition coefficients $\Gamma_{001\rightarrow mn0}$ in units of $\sqrt{\omega_r \omega_0}$, for $\omega_z/\omega_r=25$. The transitions contributing the most are the ones to low energy radial states.}
 \label{fig:F001_mno_contour}
 \end{figure}

In Fig.~\ref{fig:F001_mno_contour} we observe the dependence of these decay rates on the final radial states $n$ and $m$, observing that the main contributions come from transitions towards low energy states.

\begin{figure}[tb]
\centering
\includegraphics[scale=.5]{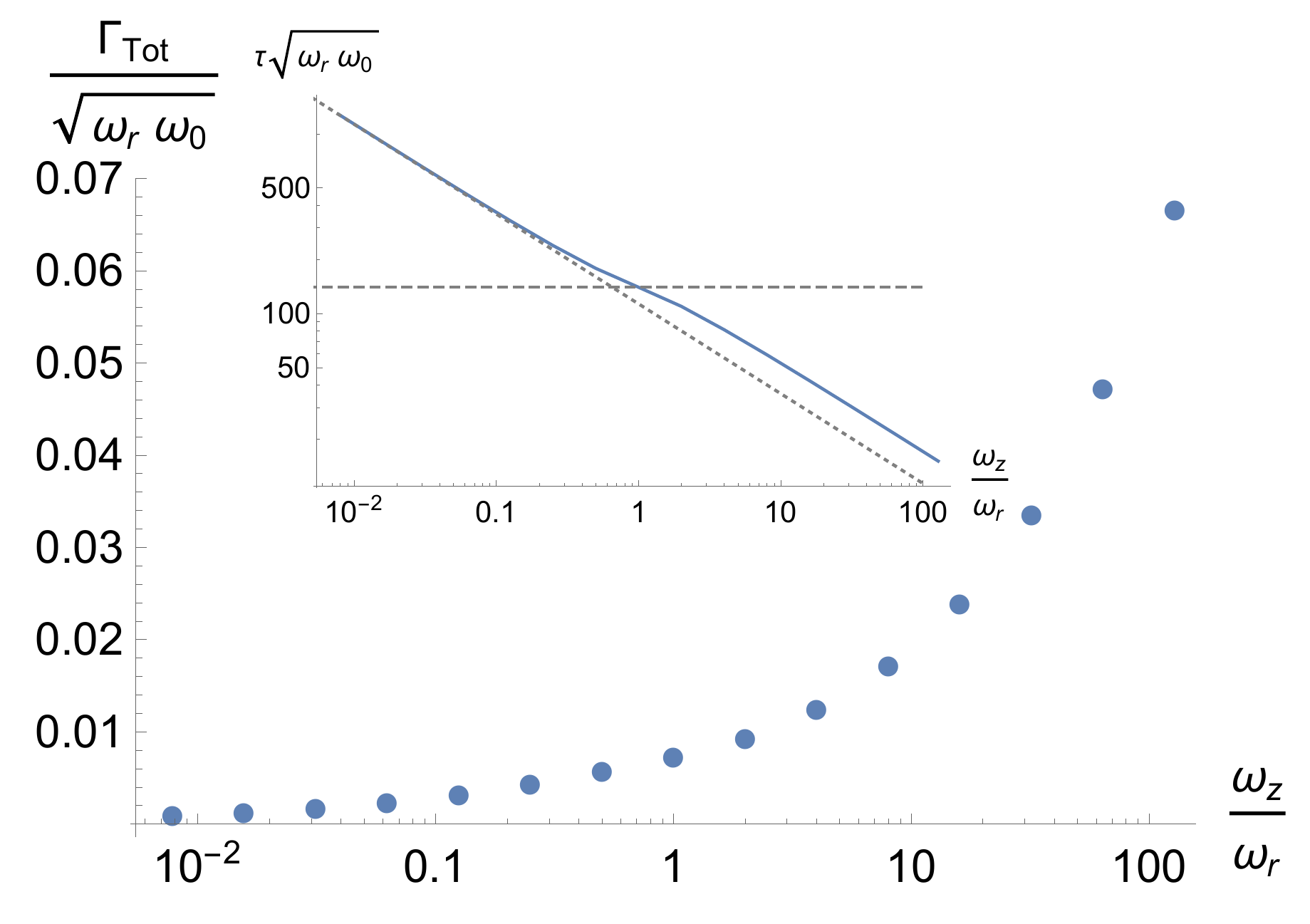} 
\caption{Total decay rate from the initial state $\ket{001}$ and decay time (inset) as a function of the ratio between the frequencies. In the 1D limit and 3D isotropic limit, respectively for $w=\omega_z/\omega_r<1$ and $w=1$, the only transition available is the one given by the decay rate $\Gamma_{001\rightarrow 000}$, for which analytical solutions in the two cases are given by Eq.~\eqref{eq:Gamma10_1D} and Eq.~\eqref{eq:Gamma10_3Diso}. The corresponding decay times are illustrated in the inset respectively with a dotted and a dashed line.
In the 1D limit, for increasing $w$, the decay rate increases due to the fact that at higher frequencies the particle will oscillate more in the same time interval, and as a consequence the interaction with the reservoir is enhanced. In the 3D limit, for $w>1$, the number of transitions contributing to the total decay rate is $w(w+1)/2$, giving an increasing total decay rate also in this limit. The value of $\omega_z$ used here is always larger than the reference frequency $\omega_0 =\mu/(2\hbar)$, so that we are always in the supersonic regime even in the low frequency 1D limit.}
\label{fig:F001_mno_vs_w}
\end{figure}

We can estimate the decay time $\tau=1/\Gamma_{Tot}$ from the initial state $\ket{0,0,1}$, after evaluating the total decay rate $\Gamma_{Tot}=\Gamma=\sum_{n,m}\Gamma_{001\rightarrow mn0}$.
In Fig.~\ref{fig:F001_mno_vs_w} we observe the variation of the total decay rate with the ratio between the two trapping frequencies  $w=\omega_z/\omega_r$ ranging through different configurations (i.e. 1D for $w<1$, 3D isotropic at $w=1$ and 3D anisotropic when $w>1$). We see that increasing the ratio $w$ in the 1D limit, the total decay rate (given by the single transition $\Gamma_{1\rightarrow 0}$) increases, as a consequence of the fact that increasing the trapping frequency along $z$ (keeping $\omega_r$ fixed) increases the number of collisions with the reservoir in a time unit.
In this limit the decay rate can be written in the simplified form \cite{DaleyPRA}
\begin{equation}
\frac{\Gamma_{1\rightarrow 0}}{\sqrt{\omega_r\omega_0}}=\frac{g_{ab}^2\rho_0\sqrt{m_a m_b}}{\pi \hbar^3 u}\sqrt{\frac{\omega_z}{\omega_r}}\int_{-\sqrt{m_b/m_a}}^{\sqrt{m_b/m_a}}e^{-\xi^2}\xi^2d\xi.
\label{eq:Gamma10_1D}
\end{equation}
In the 3D isotropic case ($w=1$), the analytical expression obtained from Eq.~\eqref{eq:F001_mno} (for $m=n=0$) can be written as
\begin{equation}
\frac{\Gamma_{001\rightarrow 000}^{iso}}{\sqrt{\omega_r\omega_0}}=\frac{2 e^{-m_b/m_a} g_{ab}^2 m_b^2 \rho_0}{3\pi u \hbar^3 m_a}.
\label{eq:Gamma10_3Diso}
\end{equation}
In the 3D limit, going towards higher values of $w>1$, although the single values of the allowed transitions $\Gamma_{001\rightarrow mno}$ decrease for increasing $w$, the total decay rate increases, since the number of available final states contributing to that is given by $w(w+1)/2$.
From the inset in Fig.~\ref{fig:F001_mno_vs_w} we see how the decay time $\tau=1/\Gamma$ varies in the different limits: the 1D asymptotic behaviour obtained from Eq.~\eqref{eq:Gamma10_1D} is represented with the dotted line and the the value obtained with the 3D isotropic analytical case of Eq.~\eqref{eq:Gamma10_3Diso} is shown for $w=1$ by the dashed horizontal line.
From the inset in Fig.~\ref{fig:F001_mno_vs_w} we see that $\Gamma_{tot}$ for the 3D case, as in the 1D case, still behaves as $\Gamma_{tot}/\sqrt{\omega_r\omega_0}\propto\sqrt{\omega_z/\omega_r}$. 

\subsection{Experimental parameters}
Let us use the above results to give an idea of the real time scales of the dynamics by using some realistic numerical parameters usually used in some dual species experiments.
In particular, we consider the case of $^{171}$Yb impurities immersed in a $^{87}$ Rb superfluid with a density $\rho_0\sim 10^{14}$ cm$^{-3}$.
Considering a scattering length $a_{bb} = 100 a_0$, with $a_0$ being the Bohr radius, we obtain that the chemical potential is $\mu_b=g_{bb}\rho_0 = \frac{4\pi\hbar^2 a_{bb}}{m_b}\rho_0\sim 3\times 10^{-11} eV$. This value of the chemical potential sets the speed of sound in the superfluid to be $u=\sqrt{\frac{\mu}{m_b}}\sim 0.5$ cm/s, and we define the reference frequency as $\omega_0 = \dfrac{\mu}{2\hbar} \sim 2\pi\times 4$ kHz.
By considering trapping frequencies $\omega_z=2\pi\times 60$ kHz $\approx 15\omega_0$ and $\omega_r = 2\pi\times 200$ Hz $\approx 0.05 \omega_0$, from the results of the previous section, we obtain that the decay time is $\tau\sim 2$ ms.

\section{Single atom cooling in a 2D harmonic trap tightly confined in one direction}
\label{Sec:2D}
In this section we use the concepts introduced above to study a different geometry for the trapping potential, where $\omega_y\gg\omega_z\gg\omega_x$.
Thinking about a cigar shaped configuration, we now call the direction along $z$ the radial one and we refer to the one along $x$ as the axial direction.

For the purposes of this study, we neglect the direction along $y$, as the states at those energies will not be involved, so effectively we study a 2D harmonic trapping with tight confinement in the direction $z$, where again we consider only the two accessible states $\ket{0}_z$ and $\ket{1}_z$.
Differently from the previous case, this time we consider the scenario where the atom can be initially excited also along the axial direction, so that we consider transitions of the kind $\ket{n_x,1_z}\rightarrow \ket{m_x,0_z}$.

In the following we estimate the transition coefficients for the aforementioned configuration and introduce some further useful approximations.
We additionally include in this section a finite temperature reservoir gas, making quantitative considerations on the effects this has in the dynamics of the impurity.

\subsection{Estimation of the transition coefficients}
Since the states involved in the $z$ direction are restricted to $\ket{0}_z$ and $\ket{1}_z$, we have two contributions to the decay of the atoms: the first given by the decay from the radially excited state, described by the decay rates $\Gamma_{n,1\rightarrow m,0}$ and the second given by the transitions from and to axial states in the same radial one, given by $\Gamma_{n,\alpha \rightarrow m,\alpha}$, with $\alpha$ either $0$ or $1$, which are effectively in 1-D.
Since we still operate in the regime $\hbar\omega_z\gg\mu_b$, for the transitions from the excited radial state we can still consider the system in the supersonic regime. For the transitions between axial states however, we have to drop this assumption, as the energy spacing $\hbar\omega_x$ in this direction can now be of the same order of the chemical potential $\mu_b$.

In Appendix~\ref{Appendix_semiclassical} we derive the decay rates $\Gamma_{n,1\rightarrow m,0}$ with an approach analogous to the one used in the previous section. However, for numerical reasons, in order to avoid divergences coming from highly oscillating terms at large $m$ and $n$ in the numerical evaluation, for the results in this section we used a semiclassical approximation \cite{migdal1977qualitative}, which we derive and compare to the fully quantum form in Appendix~\ref{Appendix_semiclassical}.

\begin{figure}[!hb]
\centering
\includegraphics[scale=.34]{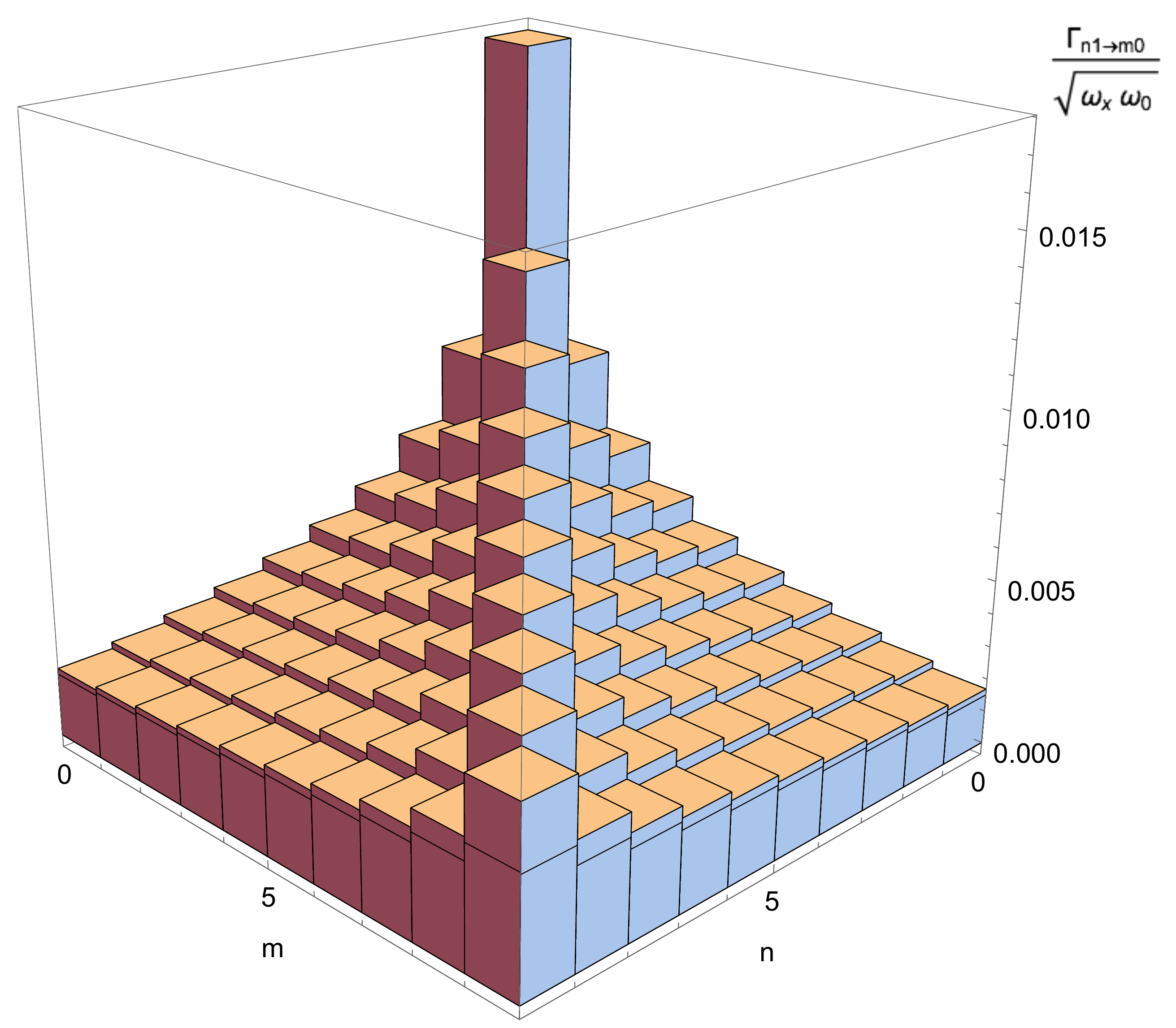}
\caption{Decay transition coefficients in 2D in units of $\sqrt{\omega_x\omega_0}$, for $\omega_z/\omega_x=100$. The transition coefficients have a maximum for equal initial and final axial states (i.e. $n=m$), but due to the constraint on the momentum coming from the energy conservation involving the ratio between the frequencies, they are not exactly symmetric about this diagonal.} 
\label{fig:Fn1_m0_contour}
\end{figure}

The decay rates between different axial and radial states (plotted in Fig.~\ref{fig:Fn1_m0_contour}), obtained by using the semi-classical approximation in the supersonic regime, are given by the expression
\begin{align}
&\Gamma_{n,1\rightarrow m,0}=\frac{2g_{ab}^2\rho_0\sqrt{m_a m_b}}{(2\pi)^2\hbar^3u}\sqrt{\frac{m_b}{m_a}(w+n-m)} \sqrt{\omega_x \omega_0} \nonumber\\
&\times\int_0^{\pi} d\theta \sin\theta B_{\phi}(n,m,\theta) J_{n-m}^2\left(\sqrt{2}\frac{x_{max}}{x_0}\xi(\theta)\right)
\label{eq:Fn1_m0}
\end{align}
where $J_{n-m}(z)$ are the first order Bessel functions. Here, 
\begin{equation}
x_{max}=x_0\left(\dfrac{\sqrt{2n+1}+\sqrt{2m+1}}{2}\right),
\label{eq:xmax}
\end{equation}
is the average between the initial and final maximum position of the impurity and where we have defined
\begin{align}
\xi^2(\theta)&=\dfrac{x_0^2k^2\cos^2\theta}{2}=\dfrac{m_b}{m_a}\left(w+n-m\right)\cos^2\theta,
\label{eq:xi_appendix}\\
\zeta^2(\theta)&=\dfrac{z_0^2k^2\sin^2\theta}{2}=\dfrac{m_b}{m_a w}(w+n-m)\sin^2\theta,
\end{align}
and
\begin{align}
B_{\phi}(n,m,\theta)&=\int_0^{2\pi}d\phi e^{-\zeta^2(\theta)\cos^2\phi}\zeta^2(\theta)\cos^2\phi
\label{eq:B_phi_appendix}\\
&=\pi\zeta^2(\theta)e^{-\zeta^2(\theta)/2}\biggl[I_0\biggl(\frac{\zeta^2(\theta)}{2}\biggr)-I_1\biggl(\frac{\zeta^2(\theta)}{2}\biggr)\biggr]\nonumber,
\end{align}
with $I_0$ and $I_1$ modified Bessel functions of the first kind.

The other contribution to the dynamics comes from the decay between radial states $\Gamma_{n,0\rightarrow m,0}=\Gamma_{n\rightarrow m}$.
As mentioned previously, however, going to high values of m makes the numerical estimation of the decay rates from and towards high states difficult, so again, as in the case of the transitions in 2D, we use the semiclassical approximation, discussed more in details in Appendix~\ref{Appendix_semiclassical}.
The general form, without assumptions on the energy of the excitations compared to the chemical potential of the BEC, is given by the expression
\begin{align}
\Gamma_{n\rightarrow m}=&\frac{g_{ab}^2\rho_0}{2\pi\hbar^2}\sqrt{\frac{m_b}{2}}\frac{\tilde{\epsilon}k^2S(k)}{\sqrt{(\tilde{\epsilon}^2+\mu_b^2)(\sqrt{\tilde{\epsilon}^2+\mu_b^2}-\mu_b})}
\label{eq:Fnm_semiclassical}\\
&\times\int_0^{\pi}J_{n-m}^2(k\cos\theta x_{max})\sin\theta d\theta, \nonumber
\end{align}
with $\tilde{\epsilon}=\hbar\omega_x(n-m)$ and where from the integration over $k$ of the delta function we obtained $k=\dfrac{\sqrt{2m_b}}{\hbar}\sqrt{\sqrt{\epsilon_k^2+\mu_b^2}-\mu_b}$.

\begin{figure}[tb]
\centering
   \includegraphics[scale=.4]{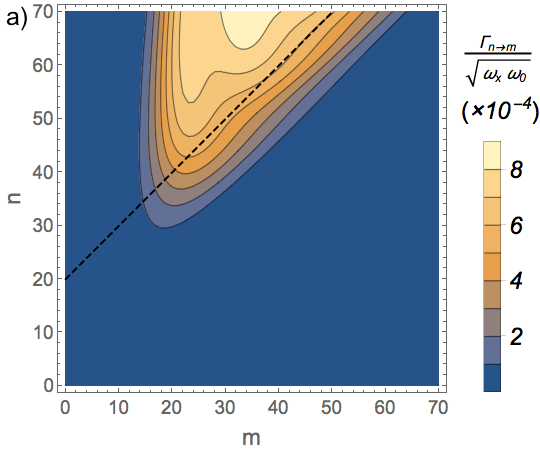} 
   \label{fig:Fnm_compared_a} 
   \includegraphics[scale=.4]{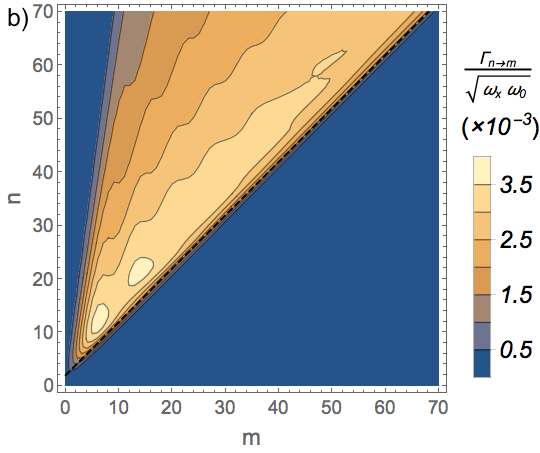}
   \label{fig:Fnm_compared_b} 
\caption{Transition coefficients $\Gamma_{n\rightarrow m}$ in units of $\sqrt{\omega_x\omega_0}$, with $\omega_x=0.1\omega_0$ (a) and $\omega_x=\omega_0$ (b), evaluated from the expression Eq.~\eqref{eq:Fnm_semiclassical} in the semiclassical approximation. The dotted black lines define the zones where $\epsilon_{\textbf{k}}=\mu_b$, i.e. in the limit between the supersonic and subsonic regimes, corresponding respectively to the areas far above and below the line.}
  \label{fig:Fnm_compared}
\end{figure}

In Fig.~\ref{fig:Fnm_compared} we show the results obtained with the most general form, Eq.~\eqref{eq:Fnm_semiclassical}, (valid for both the supersonic and subsonic limits), evaluated in the semiclassical approximation for different values of $\omega_x$. In Appendix~\ref{Appendix_semiclassical} we compare these results with the ones obtained in the fully quantum limit and show how this approximation works reasonably well even beyond the condition $|n-m|\ll n$. More precisely, the relative difference between the values obtained with the two methods is smaller than $18\%$ for $|n-m|/n\leq 0.9$, and it is in the range $0-38\%$ for transitions to and from low energy states in the limit $|n-m|\simeq n$, as discussed more in detail in Appendix~\ref{Appendix_semiclassical}.

In the following part we focus on this dynamics along one direction to study finite temperature effects that can change the dynamics and steady state of the impurity, comparing the above decay rates with the stimulated transitions rates.

\subsection{Finite temperature reservoir}
In this section we study the effects of a finite temperature reservoir, where the thermal excitations coming from the bath can, depending on the values of the temperature and chemical potential of the superfluid and the frequencies of the trap, excite the atoms either radially or axially, hence inducing reheating and changing the dynamics.
We consider the thermal energy to be always smaller than the energy scale in the radial direction ($k_BT_b\ll\hbar\omega_z$), so that the radial reheating, where the atoms would be re-excited to the first excited state along $z$, can be neglected, and we focus our analysis on two possible scenarios: $\mu_b, k_BT_b\leq \hbar\omega_x \ll \hbar\omega_z$ and $\hbar\omega_x\leq k_BT_b\leq \mu_b\ll \hbar\omega_z$.
In order to explore these regimes, we change the axial frequency $\omega_x$ and the temperature of the reservoir, keeping $\omega_z/\omega_x$ and the chemical potential $\mu_b$ fixed, as we use the frequency $\omega_0=\mu_b/(2\hbar)$ as a reference, and we compare the decay rates obtained in the previous section from Eq.~\eqref{eq:Fnm_semiclassical} with the transition coefficients associated to the absorption, as in Eq.~\eqref{eq:Hnxny_mxmy}.
As $k_BT_b\ll\omega_z$, we can neglect any absorption processes along the radial direction, therefore in the following we compare the transition rates $\Gamma_{n\rightarrow m}$ and $H_{n,m}$ and see in what regimes the reheating effects become relevant and how they would influence the final distribution at the steady state.
For the estimation of the transition coefficients we follow the same approach used for the axial decay in Eq.~\eqref{eq:Fnm_semiclassical}, where we use the full form of the structure factor and the semi-classical approximation, so that
\begin{align}
H_{n,m}=&\frac{g_{ab}^2\rho_0}{2\pi\hbar^2}\sqrt{\frac{m_b}{2}}\frac{\tilde{\epsilon}k^2S(k)}{\sqrt{(\tilde{\epsilon}^2+\mu_b^2)(\sqrt{\tilde{\epsilon}^2+\mu_b^2}-\mu_b})}\nonumber\\
&\times\frac{1}{e^{\beta\tilde{\epsilon}}-1}\int_0^{\pi}J_{n-m}^2(k\cos\theta x_{max})\sin\theta d\theta,
\label{eq:Hnm_semiclassical}
\end{align}
where $\tilde{\epsilon}=\hbar\omega_x|n-m|$ and where by integrating the delta function we obtained $k=\frac{\sqrt{2m_b}}{\hbar}(\sqrt{\tilde{\epsilon}^2+\mu_b^2}-\mu_b)^{\frac{1}{2}}$.

\begin{figure}[!t]
\centering
\includegraphics[scale=.39]{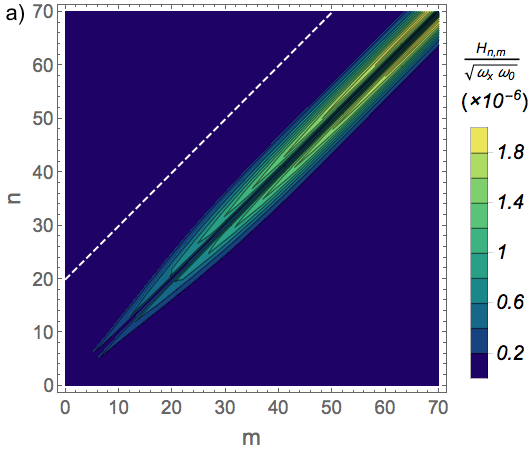}
\label{fig:Hnm_contour_compared_a}
 \includegraphics[scale=.39]{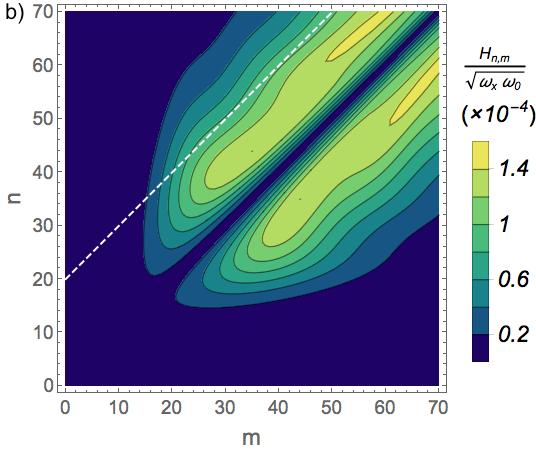}
 \label{fig:Hnm_contour_compared_b}
  \includegraphics[scale=.39]{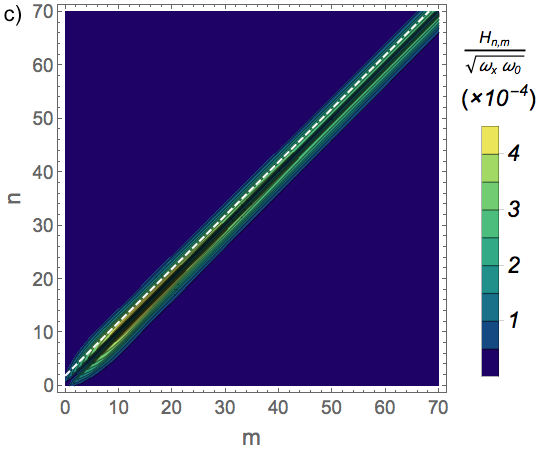}
  \label{fig:Hnm_contour_compared_c}
\caption{Transition coefficients $H_{n,m}$ in units of $\sqrt{\omega_x\omega_0}$, for different values of the bath temperature $T_b$ and trapping frequency $\omega_x$ in the two different limits $\hbar\omega_x\leq k_BT_b\leq \mu_b\ll \hbar\omega_z$ (a) and $\mu_b, k_BT_b\leq \hbar\omega_x \ll \hbar\omega_z$ (b-c). The parameters values are $T_b=0.1\hbar\omega_0/k_B, \omega_x=0.1\omega_0$ (a), $T_b=\hbar\omega_0/k_B, \omega_x=0.1\omega_0$ (b) and $T_b=\hbar\omega_0/k_B, \omega_x=\omega_0$ (c). The white dashed lines set the limit between supersonic regime (far above it) and the subsonic one (below it).}
 \label{fig:Hnm_contour_compared}
\end{figure}

In order to determine how the absorption of thermal excitations affects the dynamics of the system, we need to compare the decay rates obtained in the previous section, shown in Fig.~\ref{fig:Fnm_compared}, to the rates for the stimulated processes, represented for different values of temperature and chemical potential in Fig.~\ref{fig:Hnm_contour_compared} (as everything is in units of $\omega_0$, we equivalently vary both the temperature and the axial frequency compared to this instead, to exploit different limits).

For the case $\hbar\omega_x\leq k_BT_b\leq \mu_b\ll \hbar\omega_z$ (Fig.~\ref{fig:Hnm_contour_compared}(a)), we see that despite the thermal energy being of the same order of magnitude as the spacing between the axial energy levels, the transition coefficients are at least two orders of magnitude smaller than the decay rates of Fig.~\ref{fig:Fnm_compared}(a), so reheating effects in this regime can be neglected.
We observe how the stimulated transition coefficients change by increasing the temperature (Fig.~\ref{fig:Hnm_contour_compared}(b)) or decreasing the chemical potential (Fig.~\ref{fig:Hnm_contour_compared}(c)), moving to the limit $\mu_b,k_BT_b\leq \hbar\omega_x\ll\omega_z$.

\begin{figure}[tb]
\centering
   \includegraphics[scale=.48]{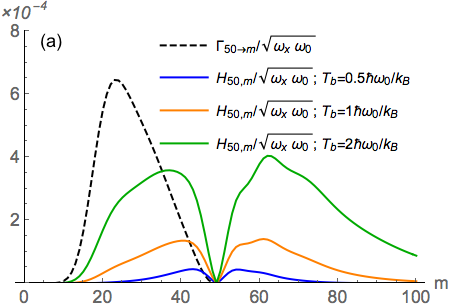} 
   \includegraphics[scale=.48]{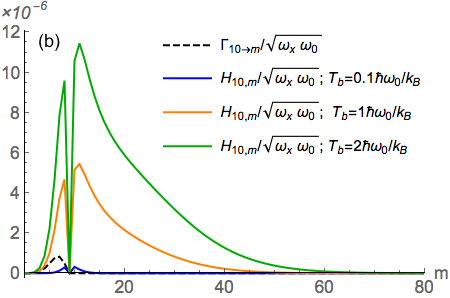} 
   \includegraphics[scale=.48]{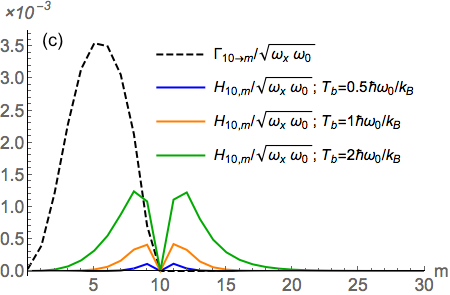} 
  \caption{Comparison between decay rates along the axial direction $\Gamma_{n\rightarrow m}$ (dashed black lines) and axial stimulated transition rates $H_{n,m}$ between the states having quantum numbers $n$ and $m$, with $n$ as shown in legends, for different values of the axial frequency $\omega_x$ and of the temperature $k_BT_b/(\hbar\omega_0)=0.5,1,2$. The values of the frequency used here are $\omega_x/\omega_0=0.1$ (a, b) and $\omega_x/\omega_0=1$ (c).}

\label{fig:Fnm_H_nm_compared}
\end{figure}

When increasing the temperature, as shown in Fig.~\ref{fig:Hnm_contour_compared}(b), not only do the stimulated transition rates increase in value, but they are also more spread towards states that are separated by a larger number of levels.
These two features are important when we compare the absorption rates with both the stimulated and spontaneous decay rates. On the one hand, if we compare these transition elements with the spontaneous decay rates in Fig.~\ref{fig:Fnm_compared}(a), we now notice that they are of the same order of magnitude. In particular, absorption from lower energy states ($n<30$) can not be neglected compared to the spontaneous decay rate.
On the other hand, the broadening of the stimulated transition coefficients about the diagonal $n=m$, as shown in Fig.~\ref{fig:Hnm_contour_compared}(b), results in absorption rates dominating over the stimulated emission rates, when considering transitions from a given initial state $\ket{n}$.
This can be better visualized in Fig.~\ref{fig:Fnm_H_nm_compared}(a-b) by comparing some specific transitions $\Gamma_{n\rightarrow m}$ and $H_{n,m}$ involving both low ($n=10$) and higher ($n=50$) energy levels, for different values of the temperature of the reservoir. Comparing the stimulated and spontaneous decay rates (for $m<n$) with the absorption decay rates ($m>n$), it is clear, especially for the case $n=10$ in Fig.~\ref{fig:Fnm_H_nm_compared}(b), that the Boltzmann distribution in the terms $H_{n,m}$, in this regime, gives absorption rates that dominate over the decay rates for both spontaneous and stimulated processes.

Conversely, if we decrease the chemical potential as in Fig.~\ref{fig:Hnm_contour_compared}(c) (this is equivalent to increasing both the temperature and frequency $\omega_x$ as we expressed them in units of $\omega_0=\mu_b/2\hbar$), the values of the stimulated rates increase respect to the case in Fig.~\ref{fig:Hnm_contour_compared}(a), but they are more narrow around the diagonal $n=m$, as the transitions involve now less states.
A comparison between these absorption rates and the spontaneous emission coefficients in Fig.~\ref{fig:Fnm_compared}(b) shows that the spontaneous emission overall prevails on the absorption for transitions for states $n\gtrsim 5$. This is due to the fact that, having lowered the chemical potential, in this case the decay will mainly be in the supersonic regime, and hence favoured by an higher value of dynamic structure factor. At the same time, as a result of the fact that the absorption processes here involve less states and the transition coefficients $H_{n,m}$ are more symmetric around $n$ even at low energy states (see Fig.~\ref{fig:Fnm_H_nm_compared}(c)), the stimulated transition coefficients already compensate the absorption rates until lower energies states at $n\approx 5$ for the given values. The combination of these effects for the emission processes, in this regime, makes the reheating effects much smaller compared to the case of Fig.~\ref{fig:Hnm_contour_compared}(a).

From the results reported in Fig.~\ref{fig:Hnm_contour_compared} and Fig.~\ref{fig:Fnm_H_nm_compared}, as discussed, we can confirm that, along with the temperature, the choice of the chemical potential of the reservoir also plays a relevant role in determining whether the absorption processes can be neglected or not. As we have seen from Fig.~\ref{fig:Hnm_contour_compared}(b), in some limits the radial reheating terms become relevant, especially for transitions from and towards lower states and can therefore affect the final configuration. We therefore studied the steady state distribution including the finite temperature effects to see how it differs when considering these contributions. This is determined by using the detailed balance condition, which can be evaluated as
\begin{align}
\bar{p}_{n+1}&=\dfrac{H_{n+1,n}}{F_{n+1\rightarrow n}+H_{n+1,n}}\bar{p}_n\\
&=\frac{H_{1,0}}{F_{1\rightarrow 0}+H_{1,0}}\times...\times\frac{H_{n-1,n-2}}{F_{n-1\rightarrow n-2}+H_{n-1,n-2}}\bar{p}_0,\nonumber
\end{align}
with $p_0=1-e^{-\beta\hbar\omega_r}$. Even though the stimulated process terms $H_{n,m}$ contain the Boltzmann distribution term coming from the number of thermal excitations in the reservoir, the distribution of probability in different states differs from the Boltzmann distribution, due to the fact that both the decay and reheating terms are affected by the structure factor.

\section{Induced dynamics of fermions in a 2D anisotropic trap}
\label{Sec:fermions}
In this section we now study the dynamics of many spin-polarized non-interacting fermions in an anisotropic harmonic trap, again in the cigar-shaped configuration $\omega_y\gg\omega_z\gg\omega_x$.
This is motivated by experiments with fermionic atoms in an optical lattice along one direction.
We start with a Fermi distribution of particles in the ground state of the harmonic oscillator along the tightly confined radial direction $z$ (i.e. single particles in the states $\ket{n_x,0_z}$), we then appropriately excite them to the first excited state along $z$ (to the states $\ket{n_x,1_z}$) and study the decay back to the ground state of $z$ and towards other states along $x$ ($\ket{m_x,0_z}$).
Since only the dynamics in two directions is involved in these processes, we treat the system in an effective 2D harmonic trap.

We determine the initial distribution of $N$ atoms at temperature $T_a$ in the radial directions given by the Fermi distribution \cite{Greiner}
\begin{equation}
\bar{n}(\epsilon_n)=\frac{1}{\exp[\beta_a\epsilon_n-\mu_a]+1},
\label{Fermi_distribution}
\end{equation}
where $\beta_a=(k_BT_{a})^{-1}$, $\epsilon_n=\hbar\omega_x n_x$ is the energy of the $n$-th excited state of the quantum harmonic oscillator (having set the zero of the energy at $\hbar\omega_x/2$) along the radial direction and in the axial ground state, and
\begin{equation}
\mu_a = \frac{\log[e^{\beta_a\epsilon_F}-1]}{\beta_a},
\end{equation}
is the chemical potential, derived by imposing the identity
\begin{equation}
N = \int_0^{\epsilon_F}g(\epsilon)d\epsilon=\int_0^{\infty}\bar{n}(\epsilon)g(\epsilon)d\epsilon,
\end{equation}
where $\epsilon_F=N\hbar\omega_x$ is the Fermi energy and $g(\epsilon)=(\hbar\omega_x)^{-1}$ is the density of states.
Considering some typical experimental values, such as $N=10^4$, $\omega_x=2\pi\times 200$ Hz and $T_a\sim 10^{-9}$ K, used for optical lattices in one dimension, we obtain $T_F=N\hbar\omega_x/k_B\sim 2\times 10 ^{-6}$ K$\gg T_a$.
This means that we can still limit our analysis to the case where, for $N$ particles, all the lower $N$ states are initially occupied, so where $\mu\rightarrow\epsilon_F$ and, under the assumption that we can excite the particles only along the axial direction resonantly with the energy $\hbar\omega_z$, the distribution of the particles in the radial states will be left invariant.
The occupation probabilities derived in Eq.~\eqref{eq:occ_prob_3D} can again be used in this case, after readapting them for the 2D scenario, so that
\begin{align}
\dot{p}_{m_x,m_z}=&\sum_{\substack{n_x>\alpha\\ n_z\geq m_z}}\Gamma_{n_x,n_z\rightarrow m_x,m_z}p_{n_x,n_z}\nonumber\\
&-\sum_{\substack{m_x'<\alpha'\\ m_z'\leq m_z}}\Gamma_{m_x,m_z\rightarrow m'_x,m'_z}p_{m_x,m_z}\nonumber\\
&+\sum_{n_x,n_z}H_{n_x,n_z;m_x,m_z}(p_{n_x,n_z}-p_{m_x,m_z}).
\end{align}
with $\alpha=m_x-\frac{\omega_z}{\omega_x}(n_z-m_z)$, $\alpha'=m_x+\frac{\omega_z}{\omega_x}(m_z-m_z')$.
Since we are dealing with non-interacting fermions, we used a stochastic description given by the Quantum Boltzmann Master Equation (QBME) \cite{Jaksch1997}, derived by neglecting the coherences in the density matrix, which leads to the following forms of the transition rates:
\begin{align}
\Gamma_{n_x,n_z\rightarrow m_x,m_z}=&\frac{2\pi}{\hbar}\sum_{\textbf{k}}|T_{n_x,n_z;m_x,m_z}(\textbf{k})|^2\delta(\epsilon_f-\epsilon_i-\epsilon_\textbf{k})\nonumber\\
&\times\bar{n}(\epsilon_i)(1-\bar{n}(\epsilon_f))
\end{align}
\begin{align}
H_{n_x,n_z;m_x,m_z}=&\frac{2\pi}{\hbar}\sum_{\textbf{k}}N(\textbf{k})|T_{n_x,n_z;m_x,m_z}(\textbf{k})|^2\nonumber\\
&\times\delta(| \epsilon_i-\epsilon_f|-\epsilon_\textbf{k})\bar{n}(\epsilon_i)(1-\bar{n}(\epsilon_f)),
\label{eq:Hnxny_mxmy}
\end{align}
where the statistics of the particles (fermions in our case) is explicitly accounted for in the terms $(1-\bar{n}(\epsilon_f))$, being $\bar{n}(\epsilon_i)$ and $\bar{n}(\epsilon_f)$ the occupation numbers respectively for the initial and final single particle energies, given by the Fermi distribution Eq.~\ref{Fermi_distribution}. We simulated the dynamics of the particles using Monte Carlo methods with jump operators \cite{Daley2014} to reconstruct the final distribution, where the advantage given by the QBME is to automatically forbid the transitions from single particle non occupied states and towards already occupied ones.

\begin{figure}[tb]
\centering
\includegraphics[scale=.5]{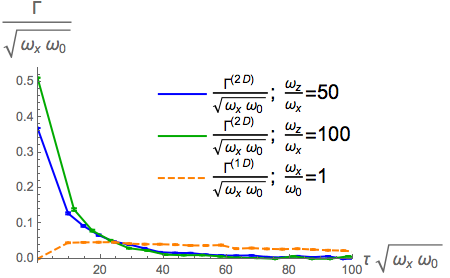} 
\caption{Total decay rates from the excited radial state $\Gamma^{(2D)}$ (solid lines) for different values of the ratio between the two frequencies and total decay rate between axial states from the ground radial direction $\Gamma^{(2D)}$ (dashed line) and for $N=8$ atoms.
The emergence of a fast and a slow decay in the two different dimensions can be seen clearly.}
\label{fig:Gamma_tot_2D_1D_vs_t}
\end{figure}

Given an initial distribution with a defined number of particles initially in the first excited state along $z$, we looked at the average number of jumps to the ground radial state in time over different repetitions, and observed that the time for all the particles to decay from the excited axial state increases when going to lower ratios $\omega_z/\omega_x$, as a consequence of the fact that the total decay rate in 2D decreases with the ratio between the axial and radial frequencies (see Fig.~\ref{fig:Gamma_tot_2D_1D_vs_t}).

The decay rate for the transitions from the particles excited along $z$ were obtained by summing over all the initially occupied and possible final states as
\begin{equation}
\Gamma^{(2D)}=\sum_{n,m}\Gamma_{n,1\rightarrow m,0},
\label{eq:Gamma_2D}
\end{equation}
while the one for radial transitions is given by
\begin{equation}
\Gamma^{(1D)}=\sum_{n',m'}\Gamma_{n'\rightarrow m'}.
\end{equation}
Even though the total decay rate is given by the sum of these two contributions, we observed them separately to see the contribution given by the radial decay in the dynamics.
As shown in Fig.~\ref{fig:Gamma_tot_2D_1D_vs_t}, the decay rate $\Gamma^{(2D)}$ given by the spontaneous emission of particles initially in the excited states $\ket{n,1}$ is maximum at $t=0$ when all the particles are excited (while $\Gamma^{(1D)}=0$) and decreases in time whilst the particles decay to $\ket{m,0}$. During this time, on the other side, because states $\ket{m,0}$ start being occupied, $\Gamma^{(1D)}$ starts increasing and then decreasing again as soon as the particles decay radially to lower states.

We observe that while the axial dynamics is fast, with $\Gamma^{(2D)}$ going to zero in the scale of $\tau\sqrt{\omega_x\omega_0}\simeq 80$ for $\omega_z/\omega_x=100$ and $N=8$ particles, the radial dynamics is much slower, so the steady state is approached in a much longer time. This is due to the fact that while the axial decay happens in the supersonic regime where the structure factor has its maximum value ($S(k)=1$), the decay rates for the radial transitions are lower even as effect of the lower structure factor that tends to suppress them.
While the decay rate in 2D, $\Gamma^{(2D)}$ [Eq.~\eqref{eq:Gamma_2D}] in these units does not depend on the choice of $\omega_x$ but only on the ratio $w$, the whole dynamics does depend on the choice of the axial frequency because this will be determined at longer times by the transitions to other axial states in 1D. For the values of the parameters used here, the 1D dynamics in the axial direction becomes dominant from $\tau\sqrt{\omega_x\omega_0}\simeq 20$, where the transition coefficients of the decays in the two different dimensions become comparable.

\begin{figure}[tb]
\centering
\includegraphics[scale=.45]{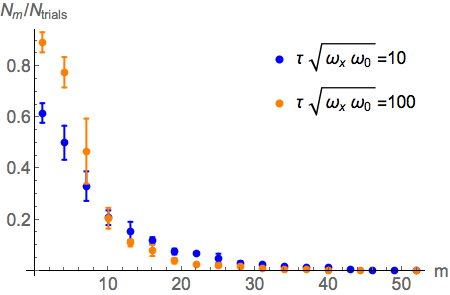} 
\caption{Occupation number of the axial modes $m$, averaged over $N_{trials}=1000$ runs of 8 atoms, at different times as shown in the legend, having set $w=100$ and $\omega_x=\omega_0$. No Pauli blockade is observed initially while the dynamics in 1D is still not dominant (see Fig.~\ref{fig:Gamma_tot_2D_1D_vs_t}), but it starts appearing when the decay between axial modes become more significant. The stationary state is reached for longer times than the one showed in the plot, as the decay rate in 1D approaches the zero more slowly.}
\label{fig:distribution}
\end{figure}

As a consequence of this, for the same parameters used in Fig.~\ref{fig:Gamma_tot_2D_1D_vs_t}, in Fig.~\ref{fig:distribution} we show the effect that the two kinds of dynamics have on the distribution of the atoms along the axial states $\ket{m,0}$. In particular, it is possible to see that for earlier times (e.g. $\tau\sqrt{\omega_x\omega_0}$, when the slow dynamics along the axial direction is not dominant yet, as compared to Fig.~\ref{fig:Gamma_tot_2D_1D_vs_t}), there is no significant effect of Pauli blocking given by the statistics of the impurities, as this starts appearing only at later times when the slower axial dynamics brings the system to the lowest energy state.

\section{Summary and outlook}
\label{Sec:conclusions}
We evaluated the decay rates of of the motional state of spin-polarized fermions immersed in a BEC and harmonically trapped in different configurations (cigar-shaped and pancake-shaped). For typical experimental parameters we estimate decay times of the order of ms, comparable to other dynamical timescales in optical lattice systems and much shorter than typical coherence timescales in experiments. We observed how the geometry of the trapping potential and the chemical potential strongly influence the decay rates and, considering a finite temperature reservoir, we showed how a convenient choice of the chemical potential can minimize the absorption of the thermal excitations for a finite temperature reservoir. Considering multiple particles towards experiments in a one-dimensional lattice of pancakes, we studied, using QBME and Monte Carlo methods, the decay of non-interacting impurities in a cigar-shaped potential and observed that the dynamics is determined by a combination of fast and a slow decay in the radial and axial directions, respectively.

This study offers some useful tools for the analytical and numerical solution of spontaneous emission of a trapped impurity in a BEC, but also for the implementation of sympathetic cooling of impurity atoms in the context of dual species experiments. We showed throughout that a semi-classical approximation is very helpful for the  estimation of transition coefficients, in regimes where rapid oscillations makes direct numerical evaluation difficult. 

This system opens possibilities as a tool for dissipative state engineering \cite{Diehl2008, Kraus2008, Verstraete2009, Yi2012}, but is also a promising environment in which to study non-Markovian open quantum systems. Indeed, the high control of the parameters of the reservoir would make it possible to explore regimes where the Markov approximation that we used in this paper does not hold anymore, either by reducing the size of the BEC reservoir or changing its trapping potential in order to have edge effects leading to backflow of information. We could also change the interaction strength via Feshbach resonances (where available), in order to go towards strong interactions. As there is not an unique approach to the study of non-Markovian systems, the possibility to explore different physical limits that are experimentally realisable, makes this system a good candidate for studies of impurities in non-Markovian reservoirs.

\section*{Acknowledgements}
We thank Suzanne McEndoo, Fran\c{c}ois Damanet, Steve Rolston, Trey Porto and Artur Widera for helpful and stimulating discussions. This work was supported in part by the EPSRC Programme Grant DesOEQ (EP/P009565/1), by the EOARD via AFOSR grant number FA9550-18-1-0064, and by the European Union’s Horizon 2020 research and innovation program under  grant agreement No.~817482 PASQuanS.

\appendix
\label{Appendix}
\section{Derivation of the master equation}
\label{Appendix_master_equation}
Here we use an open quantum system description, considering the impurity atom as the system interacting with the BEC reservoir, and derive the master equation that we used to obtain the occupation probabilities of Eq.~\eqref{eq:occ_prob_3D} for the motional states of the atoms immersed in the reservoir.
In order to study the dynamics of the trapped atom(s) interacting with the BEC, we move to the interaction picture and use the Born-Markov approximation. In addition to weak coupling, we assume that the reservoir is large enough so that we can neglect finite-size effects.
We therefore use the Redfield equation
\begin{align}
\dot{\rho}_s=-\frac{1}{\hbar^2}\int_0^t dt'\text{Tr}_B[\hat{H}_{int}(t),[\hat{H}_{int}(t'),\hat{\rho}_s(t)\otimes\hat{\rho}_B]],
\end{align}
where the density matrix of the system is $\hat{\rho}_s=\hat{\rho}_s^{(x)}\otimes\hat{\rho}_s^{(y)}\otimes\hat{\rho}_s^{(z)}$, and where we assume that the coherences can be neglected, so that we can project the density operator in each directions on the diagonal and consider $\hat{\rho}_s^{(i)}=\sum_{n_i}p_{n_i}\ket{n_i}\bra{n_i}$, with $i=x,y,z$.
In the interaction picture the operators $\hat{b}_{\textbf{k}}$ and $\hat{r}_i$ in Eq.~\eqref{eq:H_int} are now 
$\hat{b}_{\textbf{k}}(t)=e^{-\frac{i}{\hbar}\epsilon_{\textbf{k}}t}\hat{b}_{\textbf{k}}$ and $\hat{r}_i(t)=\sqrt{\frac{\hbar}{2 m_a\omega_i}(\hat{a}_i(t)+\hat{a}_i^{\dagger}(t))}$, with $\hat{a}_i(t)=e^{-i\omega_i t}\hat{a}_i$, being $i$ the index for the different spatial directions, which can be factorized. The master equation then takes the form
\begin{align}
\dot{\rho}_S&=-\frac{2 g_{ab}^2\rho_0}{V\hbar^2}\sum_{\textbf{k}}(u_\textbf{k}+v_{\textbf{k}})^2 \int_0^t d\tau\biggl[ \prod_{i=x,y,z} \sum_{n_i,m_i}\biggl(\ket{m_i} \nonumber \\
&\braket{m_i|e^{-i k_i r_i}|n_i}\braket{n_i|e^{ik_i r_i}|m_i}\bra{m_i}p_{m_i} e^{i\omega_i\tau(m_i-n_i)}\biggr) \nonumber \\
&-\prod_{i=x,y,z} \sum_{n_i,m_i}\bigl(\ket{m_i} \braket{m_i|e^{ik_i r_i}|n_i}\braket{n_i|e^{-ik_i r_i}|m_i}\bra{m_i}\nonumber\\
& p_{n_i} e^{i\omega_i\tau(n_i-m_i)}\bigr) \biggr] \left(e^{-i\epsilon_k\tau/\hbar}\braket{\hat{b}_{\textbf{k}}\hat{b}_{\textbf{k}}^{\dagger}}_B+e^{i\epsilon_k\tau/\hbar}\braket{\hat{b}_{\textbf{k}}^{\dagger}\hat{b}_{\textbf{k}}}_B\right),
\end{align}
where we can use the Markov approximation to extend the integration limit $t\rightarrow\infty$ and remove the non locality in time. Hence we obtain
\begin{align}
&\int_0^{\infty} d\tau e^{i\tau(\sum_j \omega_j(n_j-m_j)-\epsilon_{\textbf{k}}/\hbar})\\
&=\pi\hbar \delta\biggl(\hbar\sum_j \omega_j (n_j-m_j)-\epsilon_{\textbf{k}}\biggr)
\end{align}
where, as before, $j$ runs on the components in the different directions.
We find that the occupation probability  $p_{m_x,m_y,m_z}$ of the impurity in the state $\ket{m_x,m_y,m_z}$ is given by
\begin{align}
\dot{p}_{m_{x,y,z}}=&\dfrac{2\pi g_{ab}^2\rho_0}{\hbar V}\sum_{\textbf{k}}(u_{\textbf{k}}+v_{\textbf{k}})^2\\
&\times\sum_{n_{x,y,z}}\biggl(\prod_i |\braket{m_i|e^{-i k_i r_i}|n_i}|^2\biggr)\nonumber\\
&\times\biggl\{\biggl[\delta\left(\sum_i \omega_i (n_i-m_i)-\epsilon_{\textbf{k}}\right)p_{n_{x,y,z}} \nonumber\\
&-\delta\left(\sum_i \omega_i (m_i-n_i)-\epsilon_{\textbf{k}}\right)p_{m_{x,y,z}}\biggr]\braket{\hat{b}_{\textbf{k}}\hat{b}_{\textbf{k}}^{\dagger}} \nonumber\\
&+\biggl[\delta\left(\sum_i \omega_i (m_i-n_i)-\epsilon_{\textbf{k}}\right)p_{n_{x,y,z}})\nonumber\\
&-\delta\left(\sum_i \omega_i (n_i-m_i)-\epsilon_{\textbf{k}}\right)p_{m_{x,y,z}}\biggr]\braket{\hat{b}_{\textbf{k}}^{\dagger}\hat{b}_{\textbf{k}}}_B\biggr\}\nonumber.
\end{align}
The terms in the trace over the bath give $\braket{\hat{b}_{\textbf{k}}^{\dagger}\hat{b}_{\textbf{k}}}_B=N(\textbf{k})$ and $\braket{\hat{b}_{\textbf{k}}\hat{b}_{\textbf{k}}^{\dagger}}=N(\textbf{k})+1$, describing respectively  the processes of absorption of Bogoliubov thermal excitations from the reservoir and stimulated and spontaneous emission, where the distribution of excitations with momentum $\textbf{k}$ is given by the Bose distribution $N(\textbf{k})=\frac{1}{e^{\beta\epsilon_\textbf{k}}-1}$.

We then rewrite the evolution of the occupation probabilities as
\begin{align}
\dot{p}_{m_{x,y,z}}=&\sum_{\substack{n_{x,y,z}:\\ \sum_i \omega_i(n_i-m_i) > 0}} \Gamma_{n_{x,y,z}\rightarrow{m_{x,y,z}}} p_{n_{x,y,z}}\\
& -\sum_{\substack{m'_{x,y,z}:\\ \sum_i \omega_i(m_i-m'_i)>0}} \Gamma_{m_{x,y,z}\rightarrow{m'_{x,y,z}}}p_{m_{x,y,z}}\nonumber\\
& + \sum_{n_{x,y,z}}H_{n_{x,y,z};m_{x,y,z}}(p_{n_{x,y,z}}-p_{m_{x,y,z}}),\nonumber
\label{eq:occ_prob_3D_appendix}
\end{align}
where the transition rates are given by the Fermi golden rule.
In particular, defining the matrix elements as
\begin{align}
T_{n,m}(\textbf{k})=g_{ab}\sqrt{\frac{\rho_0}{V}}(u_\textbf{k}+v_\textbf{k})\braket{m_x,m_y,m_z|e^{-i\textbf{k}\cdot\textbf{r}}|n_x,n_y,n_z},
\end{align}
the decay rates are given by
\begin{equation}
\Gamma_{n_{x,y,z}\rightarrow m_{x,y,z}}=\frac{2\pi}{\hbar}\sum_{\textbf{k}}\bigl|T_{n,m}(\textbf{k})\bigr|^2\delta(\tilde{\epsilon}-\epsilon_\textbf{k}),
\label{eq:decay_rate_general_appendix}
\end{equation}
where $\tilde{\epsilon}=\delta(\hbar\sum_j \omega_j (n_j-m_j))$ is the difference of energy between initial and final state of the impurity. The transition rates of absorption and stimulated emission, used in Eq.~\eqref{eq:Hnm_semiclassical} to account for finite temperature effects, are given by
\begin{equation}
H_{n_{x,y,z};m_{x,y,z}}=\frac{2\pi}{\hbar}\sum_{\textbf{k}}N(\textbf{k})\bigl|T_{n,m}(\textbf{k})\bigr|^2\delta(\tilde{\epsilon}-\epsilon_\textbf{k}).
\label{eq:stimulated_transition_rate_general_appendix}
\end{equation}

\section{Evaluation of the decay rates and semi-classical approximation}
\label{Appendix_semiclassical}
With an analogous procedure used to estimate the transition coefficients in the 3D case (Eq.~\eqref{eq:F001_mno}), we estimated the decay rates $\Gamma_{n1\rightarrow m0}$ for the 2D configuration both in the fully quantum case and using a semi-classical approximation (Eq.~\eqref{eq:Fn1_m0}), for which we provide more details in this appendix.
Based on our previous considerations, for the particular case discussed in Sec.~\ref{Sec:2D}, as the radial trap energy spacing is much larger than the chemical potential, we can still consider the system to be in the supersonic regime, where the structure factor is $S(k)=(u_k+v_k)^2=1$.
Even in this case, for the estimation of the matrix elements, we used the relation in Eq.~\eqref{eq:tr_coeff_Laguerre}. After writing the components of the momentum in the two directions as $k_x=k\cos\theta$ and $k_z=k\sin\theta\cos\phi$, and integrating over $k$ using the properties of the delta function involving the energies, we obtain the decay rates
\begin{align}
&\Gamma_{n,1\rightarrow m,0}=\frac{2g_{ab}^2\rho_0\sqrt{m_a m_b}}{(2\pi)^2\hbar^3u}\frac{m_<!}{m_>!}\sqrt{\frac{m_b}{m_a}\bigl(w+n-m\bigr)}\sqrt{\omega_x \omega_0}\nonumber\\
&\times \int_0^{\pi} d\theta B_{\phi}(n,m,\theta)e^{-\xi^2(\theta)}\xi^{2|n-m|}(\theta)\sin\theta\bigl|L_{m_<}^{|n-m|}(\xi^2(\theta))\bigr|^2,
\label{eq:decay_rates_2D_quantum}
\end{align}
where $w=\dfrac{\omega_z}{\omega_x}$, $m_<=\min(n,m)$, $m_>=\max(n,m)$,
\begin{equation}
\xi^2(\theta)=\dfrac{x_0^2k^2\cos^2\theta}{2}=\dfrac{m_b}{m_a}\left(w+n-m\right)\cos^2\theta,
\label{eq:xi}
\end{equation}
and
\begin{align}
B_{\phi}(n,m,\theta)&=\int_0^{2\pi}d\phi e^{-\zeta^2(\theta)\cos^2\phi}\zeta^2(\theta)\cos^2\phi
\label{eq:B_phi}\\
&=\pi\zeta^2(\theta)e^{-\zeta^2(\theta)/2}\biggl[I_0\biggl(\frac{\zeta^2(\theta)}{2}\biggr)-I_1\biggl(\frac{\zeta^2(\theta)}{2}\biggr)\biggr].\nonumber
\end{align}
Here $I_0$ and $I_1$ are the modified Bessel functions of the first kind, and
\begin{equation}
\zeta^2(\theta)=\dfrac{z_0^2k^2\sin^2\theta}{2}=\dfrac{m_b}{m_a w}(w+n-m)\sin^2\theta.
\end{equation}
From Eq.~\eqref{eq:decay_rates_2D_quantum} we notice that, in contrast with the previous 3D case of Eq.~\eqref{eq:F001_mno}, the transition coefficients now contain Laguerre polynomials $L_{m_<}^{|n-m|}(x)$ that do not depend only on the difference between initial and final radial quantum numbers, but also on the particular value of $m_<$. This will make them oscillate rapidly for high values of $m_<$, giving rise to some difficulty in their numerical evaluation. In order to circumvent this problem and also optimise the time needed for their numerical evaluation, we make use of the semiclassical approximation \cite{migdal1977qualitative, DaleyPRA}, describing the motion of the impurity in the trap with a classical trajectory, so that $k_x x=k_x x_{max}\cos(\omega_x t)$ and the matrix elements of the axial transitions are
\begin{align}
&|\braket{m|e^{-ik_xx}|n}|^2=\nonumber\\
&=\biggl|\frac{2}{T}\int_0^{T/2} e^{-i k_x x_{max}\cos(\omega_xt)} \cos\left({\frac{2\pi(n-m)t}{T}}\right)dt\biggr|^2\nonumber\\
&=\biggl|\frac{\omega_x}{2\pi}\int_0^{2\pi/\omega_x}e^{-i k_x x_{max} \cos(\omega_xt)}e^{-i\omega_x(n-m)t}dt\biggr|^2\nonumber\\
&=J_{n-m}^2(k_x x_{max})
\label{eq:semiclassical_approx}
\end{align}
where $J_{n-m}(z)$ are the first order Bessel functions, 
\begin{equation}
x_{max}=x_0\left(\dfrac{\sqrt{2n+1}+\sqrt{2m+1}}{2}\right)
\label{eq:xmax_appendix}
\end{equation}
is the average between the initial and final maximum position of the impurity, and $T=2\pi/\omega_x$ is the period of the oscillations \cite{migdal1977qualitative}.
Substituting this solution for the matrix elements in the decay rates (Eq.~\eqref{eq:decay_rate_general_appendix}), we obtain the decay rates of Eq.~\eqref{eq:Fn1_m0}.

This approximation was also used for the estimation of the transition coefficients in 1D (for the slower dynamics along the axial direction), with an expression for the decay rates given by
\begin{align}
\Gamma_{n\rightarrow m}=&\frac{g_{ab}^2\rho_0}{2\pi\hbar}\int_0^{\infty}dk S(k) k^2 \delta(\hbar\omega_x(n-m)-\epsilon_k)\nonumber \\
&\times\int_0^{\pi} d\theta \sin\theta J_{n-m}^2(k\cos\theta x_{max}) d\theta \nonumber\\
=&\frac{g_{ab}^2\rho_0}{2\pi\hbar^2}\sqrt{\frac{m_b}{2}}\frac{\tilde{\epsilon}k^2S(k)}{\sqrt{(\tilde{\epsilon}^2+\mu_b^2)(\sqrt{\tilde{\epsilon}^2+\mu_b^2}-\mu_b})}\nonumber\\
&\times\int_0^{\pi}J_{n-m}^2(k\cos\theta x_{max})\sin\theta d\theta,
\label{eq:Fnm_semiclassical_appendix}
\end{align}
where in the last two lines 
\begin{equation}
k=\dfrac{\sqrt{2m_b}}{\hbar}\sqrt{\sqrt{\epsilon_k^2+\mu_b^2}-\mu_b}.
\label{eq:k_general}
\end{equation}
In order to estimate the goodness of the semi-classical approximation, we compared the decay rates obtained from this semi-classical expression and from the fully quantum one, in the 1D limit discussed here, where we can numerically evaluate them both. The quantum expression for the decay rates, in its most general form reads
\begin{align}
\Gamma_{n\rightarrow m}=&\frac{g_{ab}^2\rho_0m!}{2\pi\hbar n!}\sqrt{\frac{m_b}{2}}\frac{\tilde{\epsilon}k^2 S(k)}{\sqrt{(\tilde{\epsilon}^2+\mu_b^2)(\sqrt{\tilde{\epsilon}^2+\mu_b^2}-\mu_b})}\nonumber\\
&\times\int_0^{\pi}d\theta e^{-(x_0^2k^2\cos^2\theta)/2}\biggl(\frac{x_0^2k^2\cos^2\theta}{2}\biggr)^{n-m}\nonumber\\
&\times\biggl|L_m^{n-m}\biggl(\frac{x_0^2k^2\cos^2\theta}{2}\biggr)\biggr|^2,
\label{eq:Fnm_quantum_appendix}
\end{align}
with $k$ defined as in Eq.~\eqref{eq:k_general}.

\begin{figure}[tb]
\centering
 \includegraphics[scale=.5]{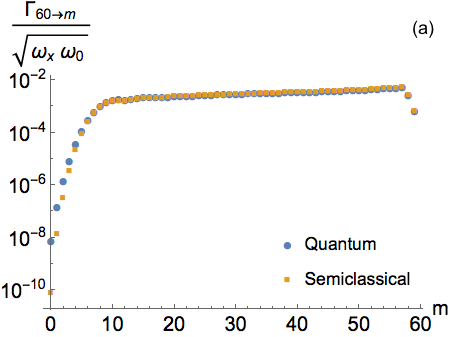}
  \includegraphics[scale=.53]{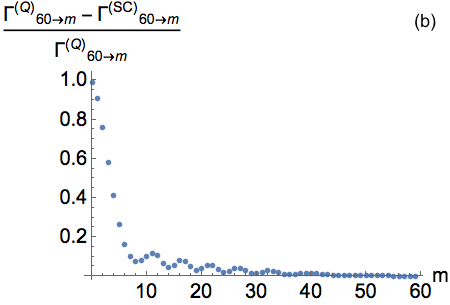}

\caption{Comparison (a) and relative difference (b) between decay rates estimated with the fully quantum expression and the semi-classical approximation, for transitions from the state $n=60$ and for a value of the trapping frequency $\omega_x=\omega_0$.}
\label{fig:quantum_vs_semicl_n60}
\end{figure}

\begin{figure}[tb]
\centering
    \includegraphics[scale=.5]{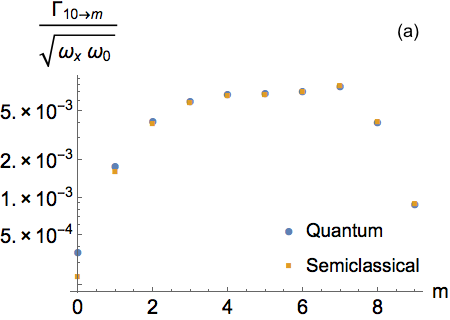}
    \includegraphics[scale=.51]{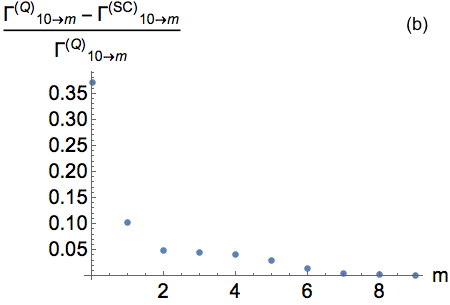}
\caption{Comparison (a) and relative difference (b) between decay rates estimated with the fully quantum expression and the semi-classical approximation, for transitions from the state $n=10$ and for a value of the trapping frequency $\omega_x=\omega_0$.}
\label{fig:quantum_vs_semicl_n10}
\end{figure}

In Fig.~\ref{fig:quantum_vs_semicl_n60} and Fig.~\ref{fig:quantum_vs_semicl_n10}  we compare the values of the decay rates obtained with both the semi-classical and fully quantum approaches, for two specific transitions from a high and from a low excited state, respectively $\Gamma_{60\rightarrow m}$ and $\Gamma_{10\rightarrow m}$. As expected from the theory, the semi-classical approximation works extremely well in the case $|n-m|\ll n$, while it is less accurate in the case $|n-m|\approx n$. More precisely, from Fig.~\ref{fig:quantum_vs_semicl_n60} and Fig.~\ref{fig:quantum_vs_semicl_n10}, we can see that the relative difference between quantum and semi-classical results are smaller than $15\%$ for $|n-m|/n \leq 0.9$.
In particular, Fig.~\ref{fig:quantum_vs_semicl_n60}(b) shows an increase of the relative difference above $40\%$ for $m\leq 4$. However, in this case, the decay rates obtained for such transitions under the semi-classical approximation and with the full quantum approach are respectively of the orders of $10^{-10}$ and $10^{-8}$ and can be approximated to 0 as they are much smaller than the other transition coefficients at higher $m$, as it can be seen from Fig.~\ref{fig:quantum_vs_semicl_n60}(b).

In general, we observe that, for transitions $n\rightarrow m$ satisfying the condition $|n-m|/n\leq 0.9$, the relative difference is below 18$\%$, and that transitions to states $m$ with $|n-m|/n>0.9$ start becoming non negligible for $n\lesssim 12$. We therefore present in Fig.~\ref{fig:quantum_vs_semicl_n10} the decay rates obtained with the two methods and the relative difference, for transitions from the initial motional state $n=10$.

We see that, while the decay rates towards small $m$ are not negligible, we have a relative difference below 10$\%$ for $m>0$ (corresponding to $|n-m|/m=0.9$), and it increases to 37$\%$ only when $|n-m|=n$. Going to even lower initial states $n$, we observed that the maximum relative difference (at $m=0$) keeps decreasing and lies within the range $0-38\%$, while for $m>0$ we still have relative discrepancy below 10$\%$. Following these considerations, we can therefore say that the semi-classical approximation has a high accuracy until $|n-m|/n=0.9$, going beyond the condition $|n-m|\ll n$ predicted by the WKB approximation for the results to be accurate, and the values obtained for other non negligible transitions with $|n-m|=n$ have a relative difference varying in the range $0-38\%$, getting smaller as the contributions from this transitions increase (at lower $n$).


\section{Evaluation of the 1D decay rates in supersonic and subsonic limits}
We can find some simplified expressions for the decay rates in 1D when considering the two limits $\epsilon_{k}\gg \mu_b$ and $\epsilon_k\ll\mu_b$ for the supersonic and subsonic regimes.
Under these conditions, they are respectively given by \citep{DaleyPRA}
\begin{align}
\Gamma_{n\rightarrow m}^{(sup)}=&\frac{g_{ab}^2\rho_0 m_b\sqrt{m_a m_b}}{\pi\hbar^4l_0}\frac{m!}{n!}\sqrt{\omega_x \omega_0}
\label{eq:Fnm_supersonic_quantum}\\
&\times\int_{-\sqrt{\frac{m_b}{m_a}(n-m)}}^{\sqrt{\frac{m_b}{m_a}(n-m)}}d\xi e^{-\xi^2}\xi^{2(n-m)}\bigl|L_m^{n-m}(\xi^2)\bigr|^2, \nonumber
\end{align}
and
\begin{align}
\Gamma_{n\rightarrow m}^{(sub)}=&\frac{g_{ab}^2\rho_0 l_0}{4\pi\hbar^2 u^4}\frac{m!}{n!}\sqrt{\frac{2m_a}{m_b}}\omega_x^2 (n-m)^2\sqrt{\omega_x \omega_0}
\label{eq:Fnm_subsonic_quantum} \\
&\times\int_{-\frac{x_0\omega_x(n-m)}{\sqrt{2}u}}^{\frac{x_0\omega_x(n-m)}{\sqrt{2}u}}d\xi e^{-\xi^2}\xi^{2(n-m)}\bigl|L_m^{n-m}(\xi^2)\bigr|^2,\nonumber
\end{align}
where $l_0=\sqrt{\dfrac{\hbar}{m_b\omega_0}}$. For transitions $\ket{1}\rightarrow{\ket{0}}$, Eq.~\eqref{eq:Fnm_supersonic_quantum} reduces to Eq.~\eqref{eq:Gamma10_1D}.

In the semi-classical approximation, under the considerations highlighted in the previous section, they respectively reduce to the two forms
\begin{align}
\Gamma_{n\rightarrow m}^{(sc-sup)}=&\frac{g_{ab}^2\rho_0 m_b \sqrt{m_a m_b}}{\pi\hbar^4l_0(\sqrt{2n+1}+\sqrt{2m+1})}\sqrt{\omega_x\omega_0}\nonumber\\
&\times\int_{-\sqrt{\frac{m_b}{2m_a}(n-m)}x_{max}}^{\sqrt{\frac{m_b}{2m_a}(n-m)}x_{max}}d\alpha J_{n-m}^2(\alpha),
\label{eq:Fnm_supersonic_semiclassical}
\end{align}
\begin{align}
\Gamma_{n\rightarrow m}^{(sc-sub)}=&\frac{g_{ab}^2\rho_0 l_0 \omega_x^2(n-m)^2}{2\pi\hbar^2u^4(\sqrt{2n+1}+\sqrt{2m+1})}\sqrt{\frac{m_a}{m_b}}\sqrt{\omega_x\omega_0}\nonumber\\
&\times\int_{-\omega_x(n-m)x_{max}/u}^{\omega_x(n-m)x_{max}/u} d\alpha J_{n-m}^2(\alpha).
\label{eq:Fnm_subsonic_semiclassical}
\end{align}

\bibliographystyle{apsrev4-1}
\bibliography{Impurities_draft_v3}


\end{document}